# An agent-based intelligent environmental monitoring system

Ioannis N. Athanasiadis and Pericles A. Mitkas

**ABSTRACT**

Fairly rapid environmental changes call for continuous surveillance and on-line decision-making. Two areas where IT technologies can be valuable. In this paper we present a multi-agent system for monitoring and assessing air-quality attributes, which uses data coming from a meteorological station. A community of software agents is assigned to monitor and validate measurements coming from several sensors, to assess air-quality, and, finally, to fire alarms to appropriate recipients, when needed. Data mining techniques have been used for adding data-driven, customized intelligence into agents. The architecture of the developed system, its domain ontology, and typical agent interactions are presented. Finally, the deployment of a real-world test case is demonstrated.

## Introduction

Environmental Information Systems (EIS) is a generic term that describes the class of systems that perform one or more of the following tasks: environmental monitoring, data storage and access, disaster description and response, environmental reporting, planning and simulation, modeling and decision-making. As the requirements for accurate and timely information in these systems are increasing, the need for incorporating advanced, intelligent features in EIS is revealed. In this context advances in Information Technology (IT) sector are promising to satisfy these requirements. Enviromatics (an abbreviation of the term "environmental informatics") is the research initiative examining the application of Information Technology in environmental research, monitoring, assessment, management, and policy (IFIP 1999).

The deployment of modern, intelligent systems for monitoring and evaluating meteorological data series and assessing air quality is only one of several enviromatics applications. Our focus in this work is to take advantage of powerful tools for software development and demonstrate their application for monitoring air-quality attributes.

Air Quality Operational Centers (AQOC) have been established worldwide in areas with potential air pollution problems. Their responsibility is to monitor critical atmospheric variables and publish regularly their analysis results. Currently, real-time decisions are made by human experts, whereas mathematical models are used for off-line study and understanding of the atmospheric phenomena involved. For example, in the London Air Quality Network (http://www.erg.kcl.ac.uk/london/) flexible data analysis is supported through statistical tools and in the Texas Natural Resource Conservation Commission (http://www.tnrcc.state.tx.us), meteorologists set the criteria for making predictions.

Several Environmental Monitoring Information Systems (EMIS) have been developed worldwide for assisting environmentalists in their tasks. These systems vary from platforms for integrating heterogeneous data sources, to real-time monitoring systems. For example, the *New Zealand Distributed Information System* (NZDIS) is an agent-based architecture for integrating environmental information systems (Purvis et al. 2003). Similarly, the *InfoSleuth project* used agents for querying distributed environmental data-clusters in a transparent way (Nodine et al. 2000) and *Java Agents for Monitoring and Management* use a collection of software agents that can perform various administrative tasks in a monitoring network (Tierney et al. 2000). Other approaches towards EMIS include the *Knowledge-based Environmental Data Analysis Assistant project*, which provides an interface between the user and off-the-shelf analysis packages for off-line study (Fine 1998) and *DNEMO*, a multi-agent system for managing air quality in real-time (Kalapanidas & Avouris 2002). These systems incorporate domain knowledge and their reasoning capabilities are deployed in the form of expert systems, case-based reasoning, or knowledge bases.

The system described in this paper, adapts a different approach since it employs data mining techniques for adding customized intelligence into an EMIS. The system, called O$_3$RTAA, is developed as a multi-agent system. Several software agents co–operate in a distributed agent society, in order to monitor both meteorological and air–pollutants attributes in an effort to evaluate air quality and, ultimately, to trigger alarms.

O$_3$RTAA relies on the *Agent Paradigm* for building intelligent software applications, while taking advantage of *Machine Learning* algorithms and *Data Mining* methodologies for extracting knowledge. The implemented system is equipped with powerful, advanced features, such as measurement validation, missing measurement estimation and custom alarm identification, while it also performs more rudimentary tasks, such as data monitoring, storage, and access.

*The approach*

The system developed, improves existing environmental monitoring systems by adding **customized intelligence** into their modules. Our approach is to couple the software agent paradigm with the knowledge discovery roadmap, in order to provide Intelligent Environmental Software Applications. This approach has been enabled by the use of Agent Academy, an integrated platform for creating intelligent agents, with training and retraining capabilities. (Agent Academy Consortium 2000)

According to the software agent paradigm (Jennings et al. 1998), agents in our system are autonomous problem-solvers that co-operate to achieve the overall goals of the system. Thus, certain system goals are assigned to specific agents. Advanced tasks, such as incoming measurement validation or custom alarm identification, are performed with the use of decision models discovered using data mining algorithms.

In general, the knowledge discovery roadmap is used for identifying useful patterns in vast volumes of data. In the case of environmental systems, there is certainly an abundance of data. In our approach, interesting patterns hidden in environmental data sets are discovered and subsequently embedded into our system, essentially adapting the problem-solving method to local conditions. As a consequence, decision-making is performed more accurately since the characteristics of the problem at hand may differ from general trends or 'rules of thumb'.

In the following sections the approach for developing an agent-based environmental monitoring system is described. The system's functionality and architecture are detailed and deployment steps are outlined.

## Functional Description

The overall goal of $O_3RTAA$ is to monitor air-quality data, including pollutants' distributions, measured by a single meteorological station. Currently, the station's field sensors measure these attributes and human experts are responsible for assessing these measurements and identifying possible alarms. Our system intervenes between the sensors and the experts and undertakes several tasks in order to assist humans in their evaluation. Besides the usual housekeeping tasks, such as the updating of the database with incoming measurements, $O_3RTAA$ is empowered with advanced features including measurement validation, estimation of missing values, and custom alarm identification. These advanced features are enabled through the exploitation of data mining techniques.

- The system operates between the field sensors and human experts in order to achieve specific goals, organized in three layers, as shown in Figure 1. The system goals form a pyramid, as the success of higher-level goals depends on the fulfillment of lower-level ones.

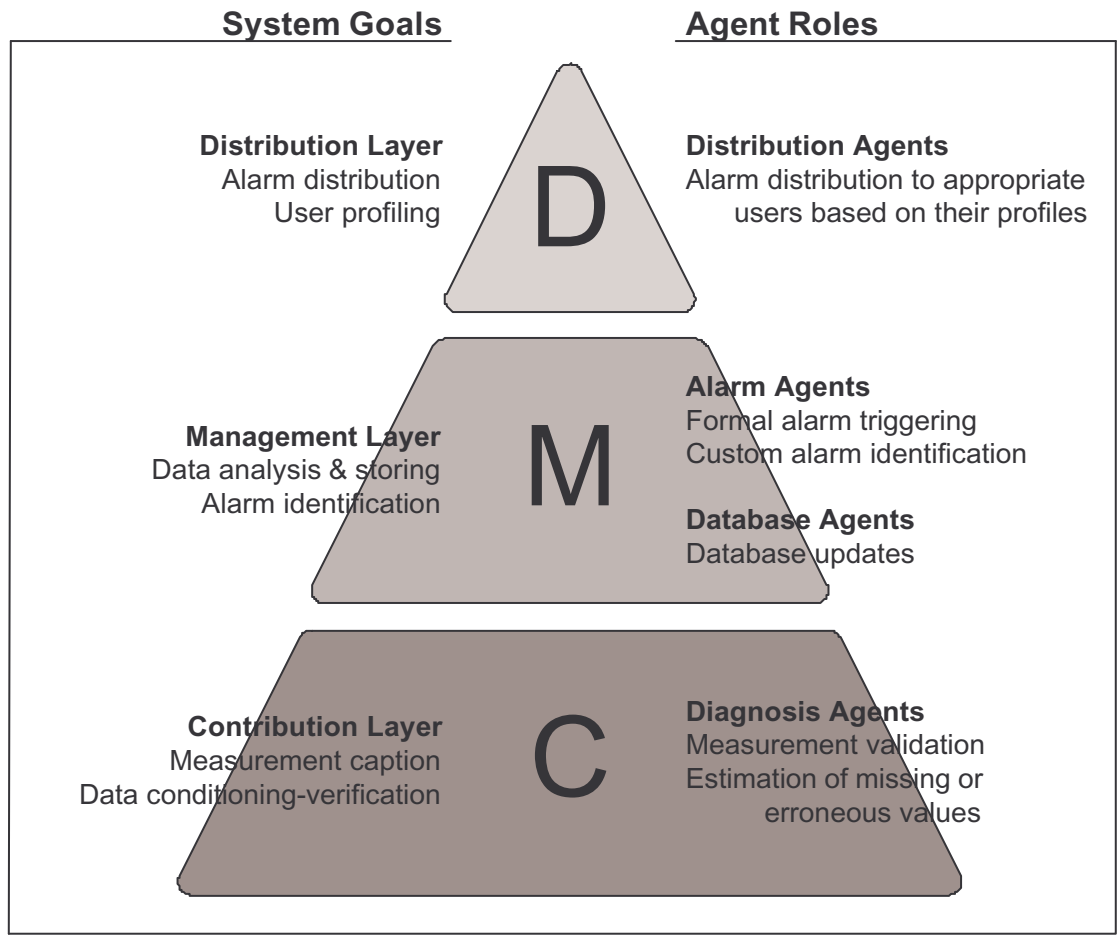

Take in Figure 1

Caption: System goals and corresponding agent roles.

At the first layer, the Contribution Layer, system goals related to preprocessing activities are gathered. Preprocessing includes efficient measurement caption, incoming data-series verification, possible malfunctions identification and restoration, and, finally, delivery of preprocessed measurement values. These tasks are assigned to the **diagnosis agent** role.

The second layer of goals, the Management Layer, is responsible for the manipulation of the preprocessed measurement values arriving from the contribution layer. Its major objective is the identification of alarms. There are two types of alarms: a) the "formal alarms", regulated by law-imposed thresholds, and b) the "custom alarms", defined by user requirements and taking under account local phenomena and trends. **Alarm agent** and **database agent** roles deliver these tasks in close collaboration.

At the top level, the Distribution Layer, goals are related to the distribution of the identified alarms to the appropriate recipients, based on previously described profiles of the interested parties. The alarms triggered at the management layer are resolved and finally delivered to the appropriate recipients at the manner they have specified (i.e., email message, SMS, etc).

## System Architecture

The implemented multi-agent system is shown in Figure 2. Agents are organized in three layers, contribution, management and distribution, following the classification of system goals discussed in the previous section. Thick arrows below the agent layers indicate the critical tasks realized through each one of them.

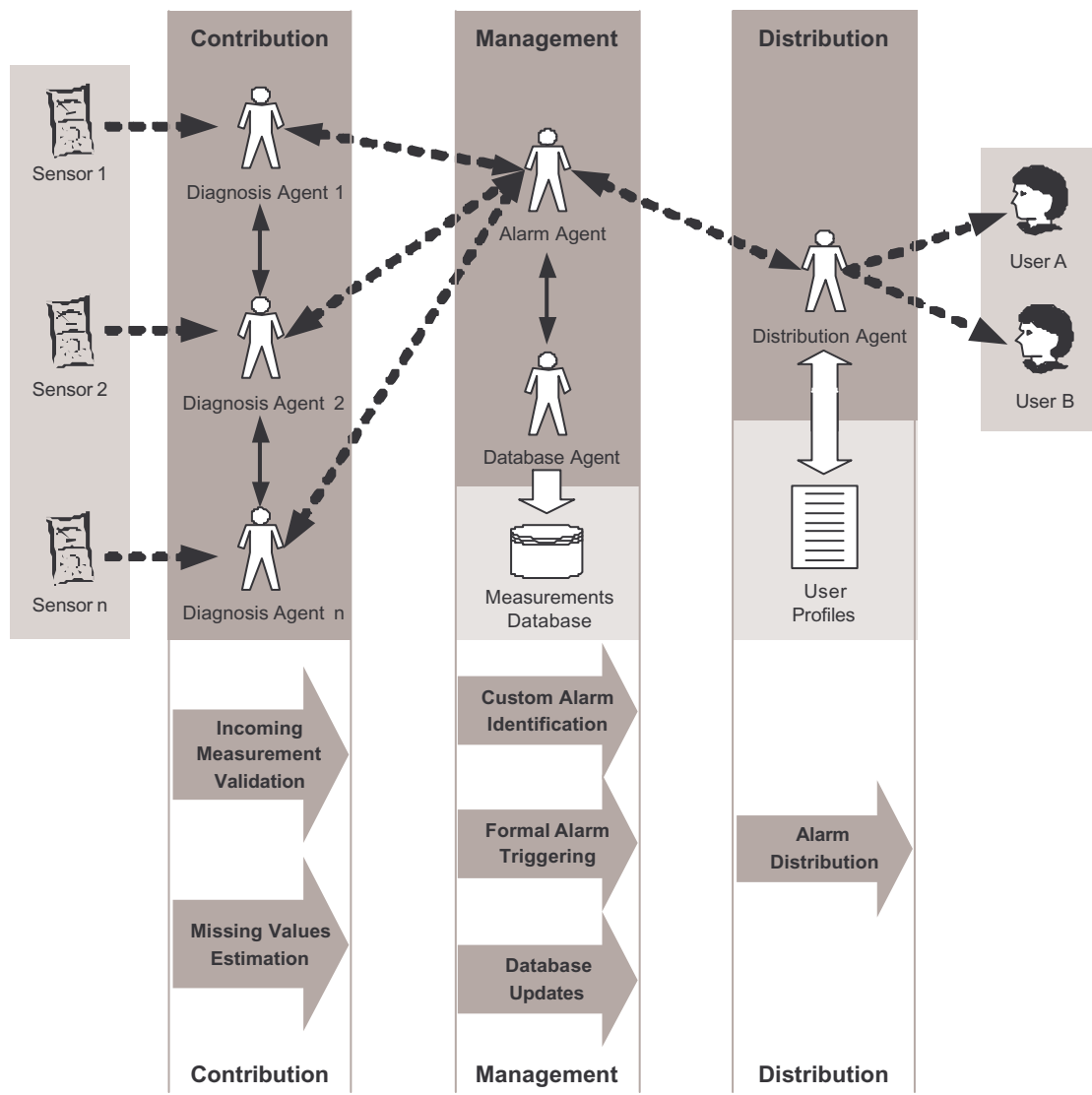

Take in Figure 2.

Caption: Agent System Architecture and related critical tasks.

Information flows from field sensors (on the left in Figure 2) to the users (on the right) through the three agent layers. Air-quality measurements arrive into the system from the sensors. Diagnosis Agents capture them, and after the validation process, deliver them to the Management layer. The Alarm Agent receives the validated measurements and determines whether a formal or custom alarm must be issued. The validated measurements are stored into the database for future use by the Database Agent. Possible alarms are forwarded to the Distribution Agent, which delivers them in the appropriate format to the corresponding end-users.

*Diagnosis Agent Type*

Diagnosis Agents are in charge of monitoring various air quality attributes including pollutants emissions and meteorological attributes. Each one of the Diagnosis Agent instances is assigned to monitor a certain field sensor. For the developed system there are eleven attributes monitored: $SO_2$ (Sulfur dioxide), $O_3$ (Ozone), NO (Nitrogen oxide), $NO_2$ (Nitrogen dioxide), $NO_X$ (Nitrogen oxides), VEL (Wind velocity), DIR (Wind direction), TEM (Temperature), HR (Relative humidity), RAD (Radiation), and PRE (Pressure). Diagnosis agents are also responsible for ensuring the efficient operation of sensors. In case of a sensor breakdown, diagnosis agents are in charge of estimating the missing values.

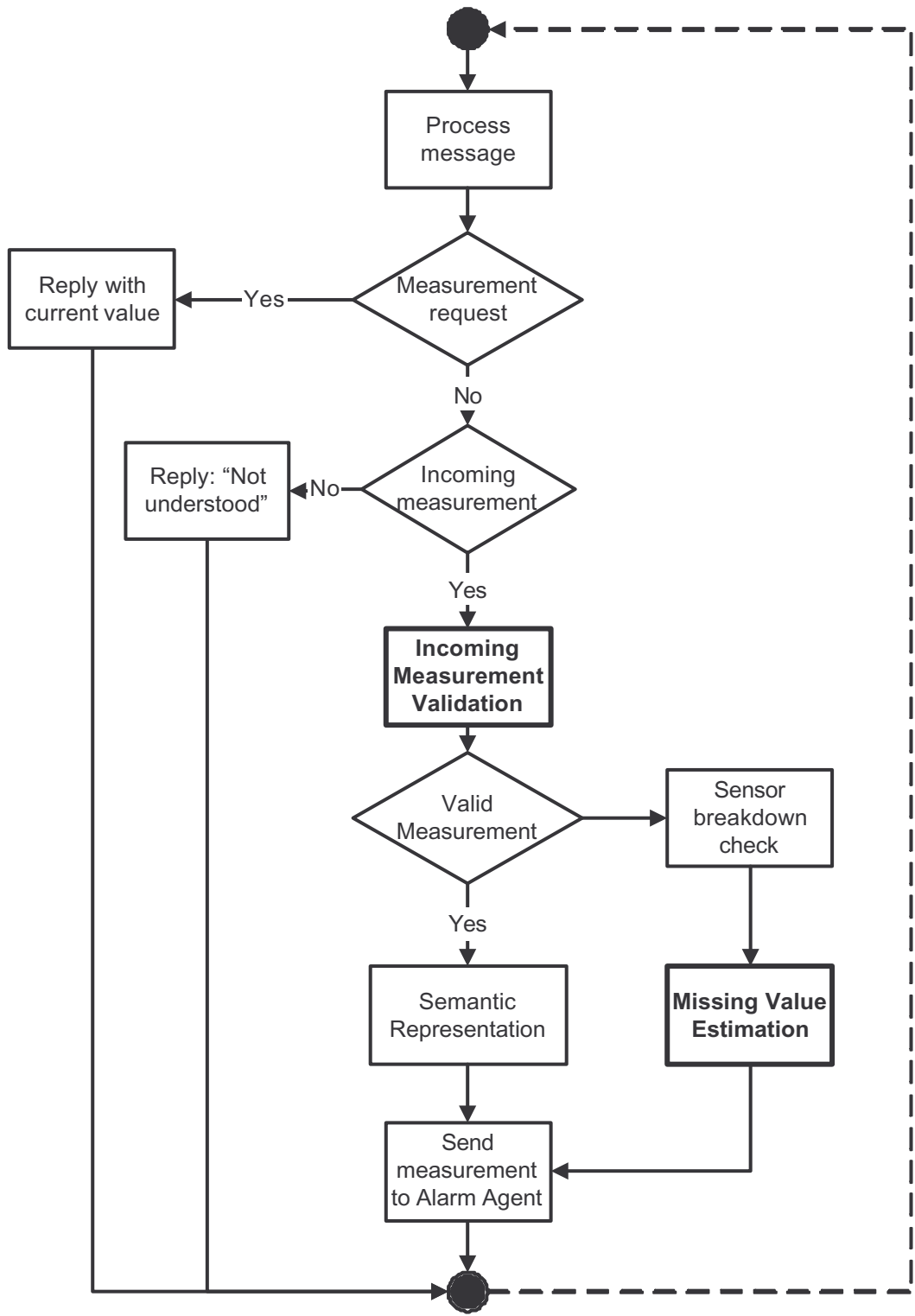

Take in Figure 3

Caption: Diagnosis Agent type internal structure. (Reasoning Engines are in bold)

The internal structure of the Diagnosis Agent type is shown in Figure 3. When a message is received by a diagnosis agent it is checked. If it is a measurement request by another agent, the diagnosis agent will respond by sending the value of its current measurement. If a new measurement has arrived, it is checked for its validity. In any other case, the incoming message is not understood and the agent responds with an appropriate message. The incoming measurement validation (IMV) procedure involves the application of the corresponding inference engine. The IMV engine runs a decision model extracted with the use of data mining on historical data. The outcome of this decision model is the classification of the incoming measurement either as 'valid' or 'invalid'.

In the first case, the valid measurement is preprocessed in a semantic way. Qualitative interpretations of the real value are calculated, such as value range, trend, or persistence. The real measurement along with the qualitative attributes is sent to the Alarm Agent.

When the measurement is classified as invalid, the Missing Value Estimation (MVE) rule engine is activated. This rule engine incorporates an inductive decision strategy discovered also by performing data mining on historical data. The MVE engine estimates the missing value level in a qualitative form (i.e. 'low value'). The qualitative indication of the missing measurement is forwarded to the Alarm Agent. When a Diagnosis Agent identifies that the sensor's measurements are persistently invalid, triggers a sensor malfunction message to the alarm agent.

The Diagnosis Agent behavior implements JADE's cyclic behavior. This is indicated in Figure 3 with the dotted arrow connecting the end state to the starting node.

*Alarm Agent Type*

Alarm Agents are responsible for triggering formal alarms and custom alarms. Formal alarms are the ones imposed by law, indicating dangerous situations in the atmosphere exceeding legal thresholds. Custom alarms are alerts for the system users about situations of their concern. In the present implementation custom alarms warn environmentalists for events based on their scientific interest related to ozone pollution. An expanded version of the system may include other type of custom alarms for patients, public administration, or industry, among others.

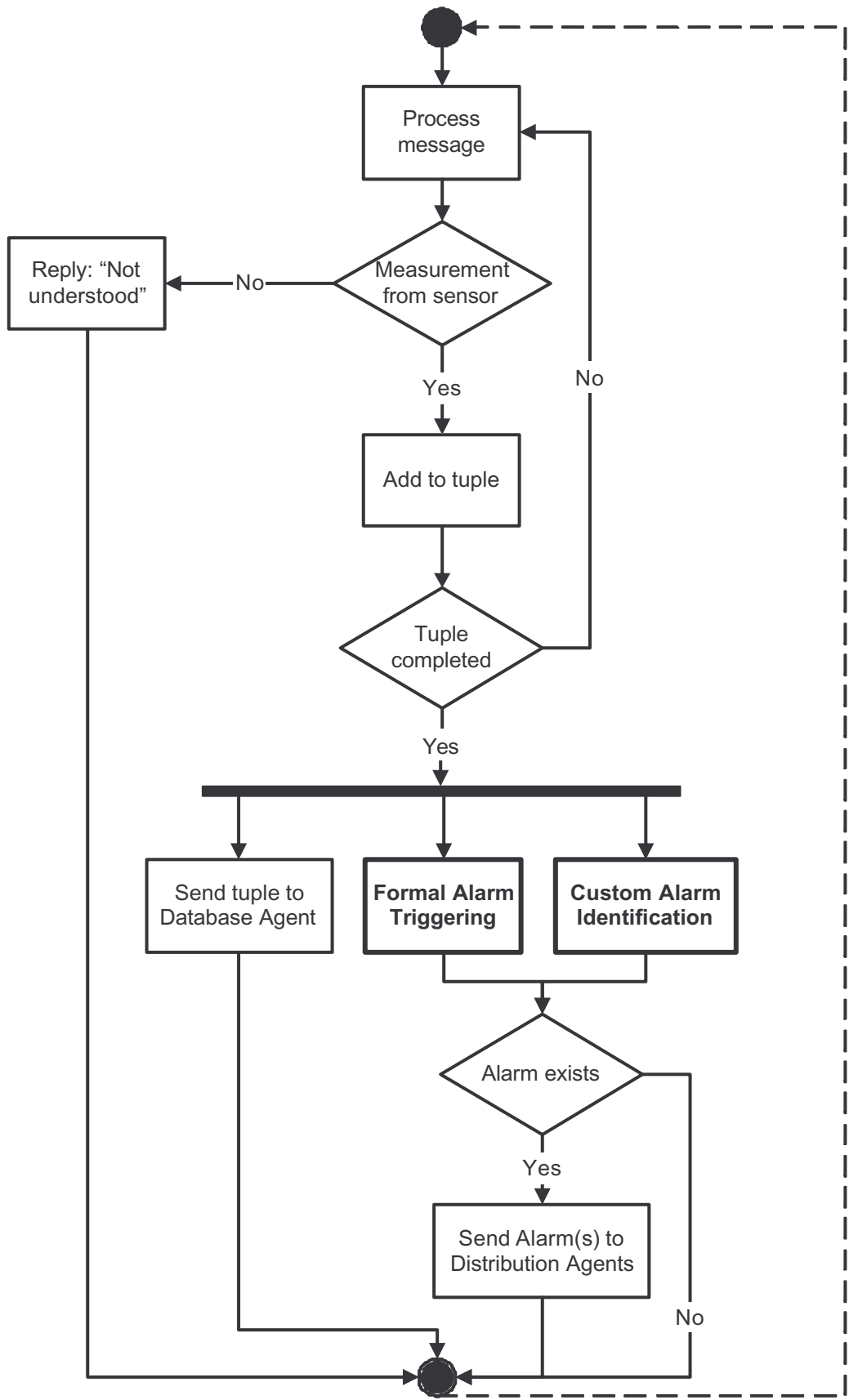

Take in Figure 4

Caption: Alarm Agent type internal structure. (Reasoning Engines are in bold)

The Alarm Agent workflow is shown in Figure 4. Alarm Agent gathers the measurements coming from field sensors through the Contribution Layer. As the Alarm Agent receives the validated measurements from the Diagnosis Agent, a tuple (record) of all concurrent measurements is created. When the tuple is completed, the Alarm Agent performs three parallel activities:

− It sends the measurements to the Database Agent in order for them to be stored to the Database.
− It may trigger 'formal alarms'. This activity involves the application of an inference engine for Formal Alarm Triggering (FAT Engine).
− It may identify and subsequently issue a custom alarm through the Identification of Custom Alarms (ICA) engine. The ICA rule engine implements a decision model extracted using data mining techniques.

Whenever Formal or Custom alarms are identified, appropriate messages are forwarded to the Distribution Agent. Note that the behavior of the Alarm Agent is also cyclic.

*Database Agent Type*

The Database Agent is responsible for updating environmental databases with data from field sensors. This task, although trivial, is vital for the system performance, as it relieves humans from manipulating all this amount of information.

The Database Agent receives a message from the Alarm Agent, containing the measurement tuple. It establishes a connection to the related databases and stores all information in the appropriate format to the corresponding table.

*Distribution Agent Type*

The Distribution Agent pushes the alarms raised by the alarm agent to the appropriate users. As an alarm message is received, the Distribution Agent queries user profiles for selecting target users interested in the alarm and selects the appropriate medium of notification (i.e. email or SMS). As the set of recipients and mediums has been prespecified, the alarm is transformed properly into an alert and finally transmitted to the users.

# System Deployment

The aforementioned agent types have been developed using the Agent Academy platform following a certain procedure discussed in Mitkas et al. 2003. This procedure involves the definition of agent type functions, domain ontology, along with agent interactions, and communication language.

*Agent Academy*

Agent Academy is a framework for developing agents empowered with training capabilities exploiting data mining techniques and is distributed under the less-GPL license. (Agent Academy Consortium 2000, Mitkas et al. 2002). Agent Academy facilitates the whole procedure for developing a multi-agent community. Agent Academy supports the creation of agents with limited initial referencing capabilities, along with a certain agent training process targeting to augment agent intelligence efficiently, through the embedding of data–driven decision strategies into them (Athanasiadis et al. 2003b).

*Technologies adopted*

The development of $O_3RTAA$ involved a variety of technologies. The JADE platform was used for agent creation (Bellifemine et al. 2000) and JESS engine for rule execution (Friedman-Hill 2003). Ontology design and specification was done with Protégé 2000 Ontology Editor (Grosso et al. 1999). Data mining experiments were performed using the core of the WEKA tool for knowledge analysis (Witten & Frank 1999) and PMML format was selected for knowledge model representation (Data Mining Group 2001). FIPA standards were adopted for agent communication (FIPA 2000).

Information flow among agents is structured in agent messages, while rule engines support data-driven decision-making.

*Agent Ontology and Messaging*

In our system, the content of agent messages is structured using a common ontology, created with the Protégé 2000 ontology editor. Part of the developed ontology is shown in Figure 5, containing the main concepts of the system. The slots of the various concepts have been configured in order to contain the appropriate information communicated by the agents. For example, concept 'Pollutants' has the following slots: value, level, and variability for describing a measurement in both quantitative and qualitative form.

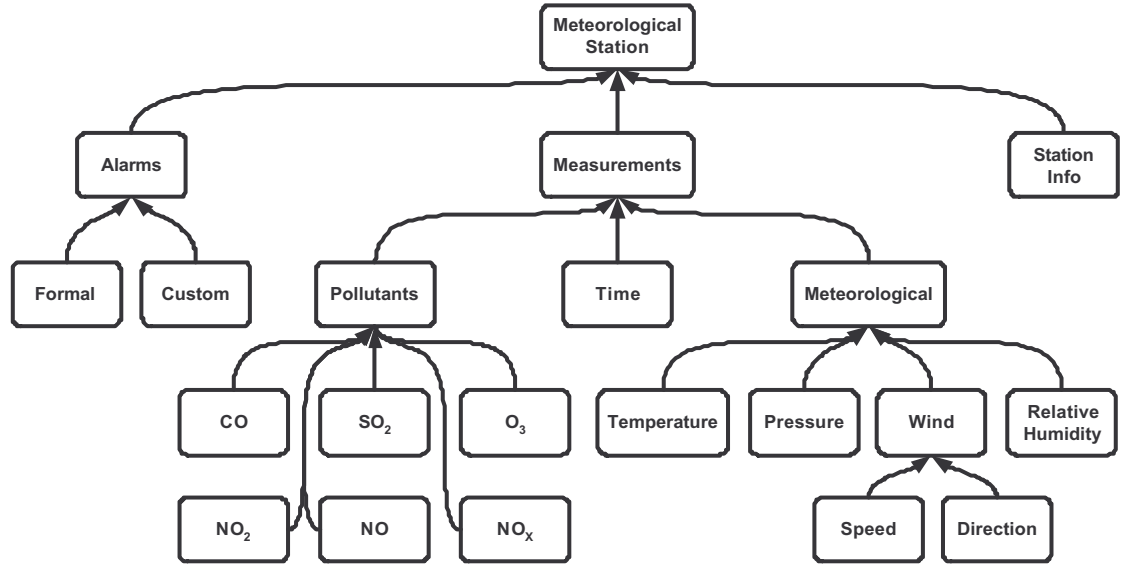

Take in Figure 5

Caption: Main concept structure as a part of the domain ontology.

Information shifts between agent layers via messages communicated among agent instances. The ontology is exported in RDFS format, which is compatible with JADE agents. The agent communication language is FIPA-ACL and a sample message between a Diagnosis Agent and the Alarm agent is shown in Figure 6. Using the implemented ontology, the ozone Diagnosis Agent reports its measurement to the alarm agent, as needed. Agent predicate concept 'sendMeasurement' is used for informing the Alarm Agent about a measurement and has two slots: one is instance of the 'TimeStamp' concept and the other an instance of the 'O3' concept. Note that O3 concept inherits its slots from the Pollutants concept.

In a similar way, all information is exchanged between agents, conforming to FIPA specifications and realizing the developed domain ontology.

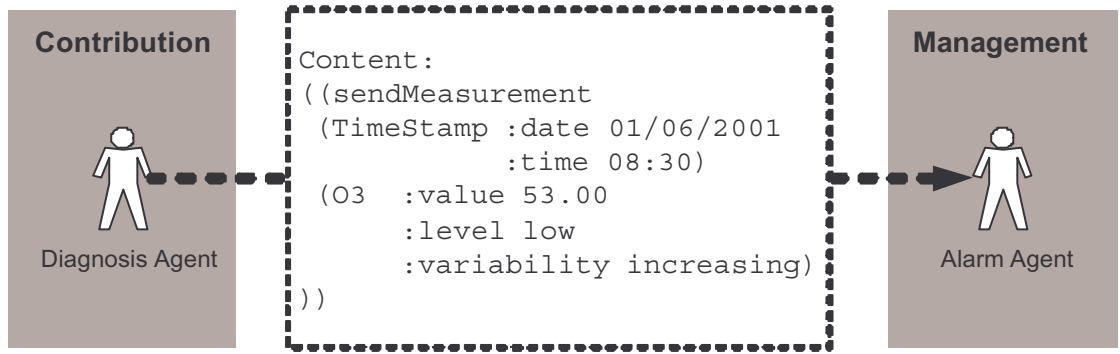

Take in Figure 6

Caption: Ozone Diagnosis Agent reports to the Alarm Agent the current measurement.

*Agent Decision Making Models*

Agent decision-making is the most critical part of the system as it is related with all advanced features of the system. O$_3$RTAA contains four different reasoning engines. IVM, MVE and ICA Engines realize inductive decision models and FAT Engine realizes a deductive decision model. Inductive decision models are those created using data mining techniques and incorporating data-driven knowledge into the system.

In general, the goal of the data mining procedure is to reveal interesting patterns from data volumes and use them for future decision making. A predictive model is built based on these patterns. Predictive models are comprised of two parts: the **predictor** and the **response**. The predictor consists of a set of independent attributes used to make the prediction, while the response is the target value or class. Predictive models could be in the form of decision trees, neural networks or association rules. A predictive model can be transformed into **logical rules** having the form: *'If assumption then consequence'* and thus is easy to implement and execute through a Rule Engine.

To develop our system, we have used data mining techniques for adding customized intelligence in the form of data-driven decision strategies. For the knowledge extraction process, Agent Academy's Data Miner component was used, which extends the Waikato Environment for Knowledge Analysis (WEKA) tool. Predictive models extracted with the Agent Academy Data Miner are formed using Predictive Model Markup Language (PMML). PMML documents are transformed into JESS rules and finally incorporated into software agents.

*The test case deployment*

Historical data coming from the Onda meteorological station, in the Community of Valencia, Spain, were used for creating data-driven rules. More specifically, several meteorological attributes and air-pollutant values, along with validation tags, were recorded on a quarter-hourly basis during years 2000 and 2001. There are about 70,000 records in the volume. Raw measurements have been preprocessed properly in order to create training and validation datasets compliant with agent goals and suitable for data mining.

As an example, let us focus on the Incoming Measurement Validation process. The dataset was preprocessed in order to contain the corresponding validation tag as the response attribute and the current value of a specific pollutant along with a set of previous values and measures as predictor attributes. These measures are shown in Table 1. The preprocessed dataset was split into two sets. Data recorded in year 2000 were used as the training set and data recorded in 2001 were used as the test set. Quinlan's C4.5 algorithm for decision tree induction (Quinlan 1993) was applied on the data. The resulted decision tree contains 15 leaves, i.e. it can be implemented as a set of fifteen rules. The predictive accuracy of the decision model on the validation set is 99.71%, which is satisfactory. Our data mining experiments are discussed in detail in Athanasiadis et al (2003a,b).

| | |
|---|---|
| O3 | The current ozone value |
| O3_30 | The ozone value 30 min ago |
| O3_90 | The ozone value 90 min ago |
| MinMax60 | The difference between the maximum and the minimum ozone value in the last 60 min |
| MinMax150 | The difference between the maximum and the minimum ozone value in the last 150 min |
| O3val | The corresponding validation tag (valid/erroneous) |

Take in Table 1

Caption: Attributes used for the validation decision model

# Summary and future work

In this paper we have presented an intelligent environmental monitoring system, developed with software agents and using data mining techniques for adding customized intelligence into them. System goals have been assigned successfully to agents, which act as mediators and deliver validated information to the appropriate stakeholders. Agent communication and semantic representation of information is performed using state of the art tools.

System customization has been based on the application of data mining techniques on historical environmental data for adding data-driven intelligence. The use of C4.5 algorithm yielded trustworthy decision models for validating incoming measurements, estimating the erroneous ones and identifying custom alarms in the described application.

The system developed will be installed as a pilot case at the Mediterranean Centre for Environmental Studies Foundation (CEAM), Valencia, Spain, in collaboration with IDI-EIKON, Valencia, Spain. Having the system validated for a single meteorological station, future work will be focused on expanding the architecture for covering the whole network.

## Acknowledgements

Authors would like to express their gratitude to the Agent Academy Consortium for their valuable help. Special thanks go to the IDI-EIKON team for their efforts within Agent Academy project to deploy the $O_3$RTAA system and to CEAM for the provision of the ONDA dataset. The Agent Academy project is partially funded by the European Commission under the IST programme (IST-2000-31050).